\documentclass{aa}  
\usepackage{graphics,graphicx}
\usepackage{xcolor}
\usepackage{enumitem}
\usepackage[dvipsnames]{xcolor}
\usepackage{txfonts}
\usepackage{longtable}
\usepackage[unicode=true,pdfusetitle,
 bookmarks=true,bookmarksnumbered=false,bookmarksopen=false,
 breaklinks=false,pdfborder={0 0 0},pdfborderstyle={},backref=false,colorlinks=true]
 {hyperref}
\hypersetup{
 urlcolor=blue,citecolor=blue}

\begin{document}
\nolinenumbers
\makeatletter
\renewcommand\makeLineNumber{} 
\makeatother
 
   \title{From Observations to Simulations: A Neural-Network Approach to Intracluster Medium Kinematics} 
    \titlerunning{Neural Network Matching of ICM Velocities} 

   \author{E. Gatuzz\inst{1}, 
           J. ZuHone\inst{2}
           J. S. Sanders\inst{1},
           A. Fabian\inst{3}, 
           A. Liu\inst{4},
           C. Pinto\inst{5}  \and
           S. Walker\inst{6}
          }   
          
   \institute{Max-Planck-Institut f\"ur extraterrestrische Physik, Gie{\ss}enbachstra{\ss}e 1, 85748 Garching, Germany\\
              \email{egatuzz@mpe.mpg.de}
         \and
             Harvard-Smithsonian Center for Astrophysics, 60 Garden Street, Cambridge, MA, 02138, USA   
         \and
             Institute of Astronomy, Madingley Road, Cambridge CB3 0HA, UK  
         \and
             Institute for Frontiers in Astronomy and Astrophysics, Beijing Normal University, Beijing 102206, China    
         \and
              INAF - IASF Palermo, Via U. La Malfa 153, I-90146 Palermo, Italy  
          \and
             Department of Physics and Astronomy, University of Alabama in Huntsville, Huntsville, AL 35899, USA                                            
             } 
   \date{Received XXX; accepted YYY} 
             

\abstract{We present a systematic comparison between {\it XMM-Newton} velocity maps of the Virgo, Centaurus, Ophiuchus and A3266 clusters and synthetic velocity maps generated from the Illustris TNG-300 simulations. 
Our goal is to constrain the physical conditions and dynamical states of the intracluster medium (ICM) through a data-driven approach. 
We employ a Siamese Convolutional Neural Network (CNN) designed to identify the most analogous simulated cluster to each observed system based on the morphology of their line-of-sight velocity maps. 
The model learns a high-dimensional similarity metric between observations and simulations, allowing us to capture subtle kinematic and structural patterns beyond traditional statistical tests. 
We find that the best-matching simulated halos reproduce the observed large-scale velocity gradients and local kinematic substructures, suggesting that the ICM motions in these clusters arise from a combination of gas sloshing, AGN feedback, and minor merger activity. 
Our results demonstrate that deep learning provides a powerful and objective framework for connecting X-ray observations to cosmological simulations, offering new insights into the dynamical evolution of galaxy clusters and the mechanisms driving turbulence and bulk flows in the hot ICM.
}

    \authorrunning{Gatuzz et al.}
   \maketitle

\section{Introduction}\label{sec_in}

The intracluster medium (ICM) is a tenuous, X-ray-emitting plasma permeating galaxy clusters. 
Despite its extremely low density and high temperatures ($\sim 10^{7}$--$10^{8}$ K), the ICM contains several times more baryons than the cluster galaxies combined \citep{and10,dai10,kra12}, making it a crucial component for understanding the physics of cosmic structures. 
Simulations predict that the ICM should host turbulent or random motions and large-scale bulk flows arising from hierarchical merging processes \citep[e.g.][]{lau09,vaz11,sch17,ha18,vaz21}. 
Additionally, sloshing motions induced by infalling substructures can generate relative bulk velocities of a few hundred km~s$^{-1}$ \citep[e.g.][]{asc06,ich19,vaz18,zuh18}. 
The central Active Galactic Nucleus (AGN) in the brightest cluster galaxy further contributes to ICM motions through jets and buoyantly rising relativistic plasma bubbles, potentially driving velocities of $\sim 100$--500 km~s$^{-1}$ in surrounding gas \citep[e.g.][]{bru05,hei10,ran15,yan16,bam19}. 

Determining the kinematic properties of the ICM is therefore essential for developing a comprehensive physical picture of galaxy clusters. 
In this context, \citet{san20} introduced an innovative technique that leverages the instrumental X-ray lines in {\it XMM-Newton} EPIC-pn spectra to calibrate the absolute energy scale with a precision better than $100$ km~s$^{-1}$ at Fe--K. 
Using this technique, direct measurements of ICM velocities have been obtained in several systems---including the Perseus and Coma clusters \citep{san20}, Virgo \citep{gat22a,gat23b}, Centaurus \citep{gat22b,gat23d}, Ophiuchus \citep{gat23a,gat23c}, and A3266 \citep{gat24a,gat24b}---and these results have been robustly confirmed by XRISM Resolve observations \citep{xri25b,xri25a,fuj25,gat25a}. 
These studies reveal a variety of velocity structures associated with both merger-driven dynamics and AGN feedback. 

\begin{figure*}
    \centering
    \includegraphics[width=0.99\linewidth]{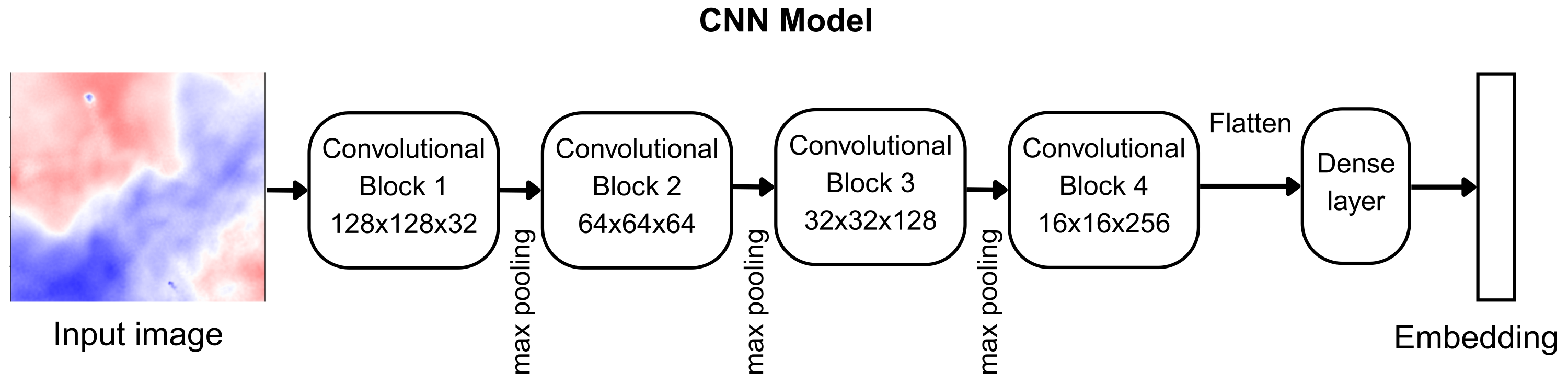}
    \caption{Diagram of the CNN model used in our analysis.
    The CNN acts as a feature extractor, converting each input velocity map into a compact embedding vector.
    Convolution and max-pooling layers progressively identify hierarchical spatial and kinematic features; the output is then flattened and passed through fully connected layers to produce the final embedding.}
    \label{fig:siam_cnn_diag_model}
\end{figure*}

The launch of the X-ray Imaging and Spectroscopy Mission ({\it XRISM}) has opened a new era of direct ICM kinematic measurements. 
Early Resolve observations reveal a wide range of dynamical states: relaxed systems can exhibit extremely low non-thermal support \citep[e.g., Abell~2029 shows $\sigma_v = 169 \pm 10$ km~s$^{-1}$ and a non-thermal pressure fraction of only $\sim 2\%$,][]{xri25a}; 
in contrast, Perseus- and Centaurus-like cool cores display measurable bulk flows consistent with sloshing, with line-of-sight velocities of $10^2$--$3 \times 10^2$ km~s$^{-1}$ detected in Centaurus \citep{xri25b,zha25}.  
XRISM measurements in Coma reveal ordered, large-scale velocity structure and modest velocity dispersions ($\sim 200$ km~s$^{-1}$ in multiple fields), implying coherent bulk flows at the few $10^2$--$10^3$ km~s$^{-1}$ level when combining several pointings \citep{gat25a}.  
Overall, early XRISM results indicate that cool-core regions generally exhibit lower velocity dispersions than predicted by several numerical models, providing stringent constraints on turbulent pressure support and AGN feedback.  
Such measurements are now becoming essential benchmarks for calibrating and testing cosmological simulations.

Comparing the observed ICM velocity structure with simulations is therefore imperative, as theoretical models predict tight correlations between velocity power spectra and thermodynamic quantities such as entropy, temperature, density, and pressure \citep{gas14,zhu14,moh19}. 
Deep-learning methods have emerged as powerful tools for bridging observations and simulations in this context. 
Convolutional Neural Networks (CNNs) have been successfully applied to tasks such as galaxy morphology classification, feature extraction in large surveys, and inference of physical properties from real and simulated data \citep{hue15,dom18,pea19}.  
Among these architectures, Siamese CNNs---networks designed to learn a similarity metric between inputs---enable robust comparison even in limited-data scenarios. 
\citet{bru19} employed a Siamese CNN for supernova classification using sparse light curves, and \citet{zha22} applied a similar architecture to galaxy morphology classification with few labeled examples, demonstrating the general utility of deep metric learning for astrophysical inference.

Here, we present a systematic comparison between the velocity maps obtained for the Virgo, Centaurus, and Ophiuchus clusters and magnetohydrodynamical simulations, using a Siamese CNN to infer the physical conditions of the ICM. 
While previous deep-learning efforts have compared other ICM observables (e.g., X-ray surface brightness or temperature) across simulations and observations, the velocity field remains the most complex and least constrained. 
We therefore focus this initial work exclusively on the kinematic structure to evaluate the performance of the Siamese framework in this challenging domain.
This paper is organized as follows. 
Section~\ref{sec_obs} describes the {\it XMM-Newton} observations used in the analysis. 
Section~\ref{sec_sim} presents the magnetohydrodynamical simulations employed. 
Section~\ref{sec_ai_stat} outlines the Siamese CNN methodology. 
Section~\ref{sec_cnn_results} reports the results of the CNN-based similarity analysis. 
Section~\ref{sec_phys} discusses and interprets the best-matching simulated velocity maps. 
Finally, Section~\ref{sec_con} summarizes our conclusions. 
Throughout this work, we adopt a $\Lambda$CDM cosmology with $\Omega_m = 0.3089$, $\Omega_\Lambda = 0.6911$, and $H_0 = 67.74~\mathrm{km~s^{-1}~Mpc^{-1}}$.

\begin{figure*}
    \centering 
    \includegraphics[width=0.75\linewidth]{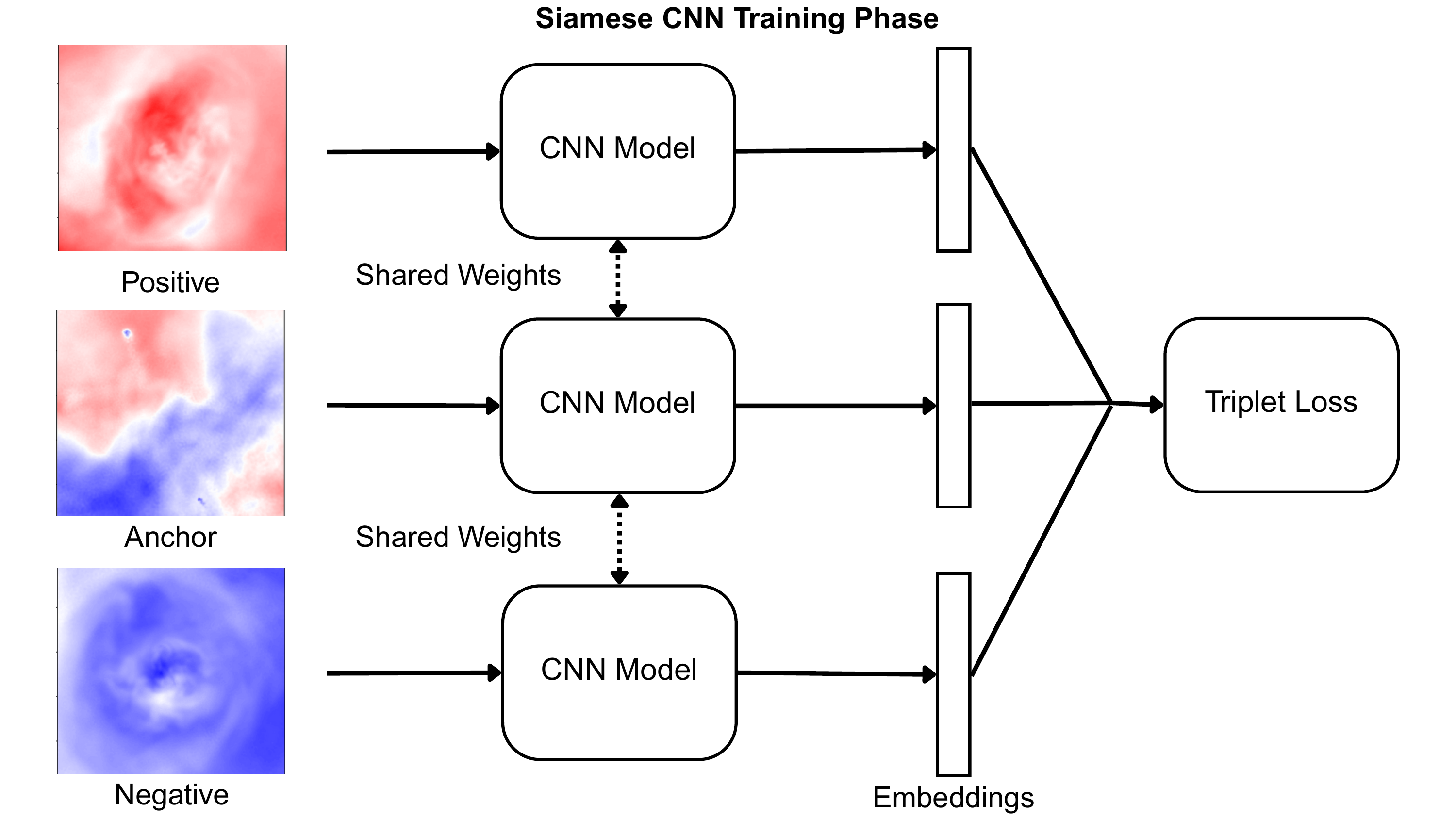} 
    \caption{Diagram of the Siamese CNN training phase.
    During training, pairs of velocity maps are passed through two identical CNN encoders with shared weights. 
Each triplet consists of a reference (Anchor), a matched projection from the same simulation (Positive), and a mismatched projection from a different simulation (Negative). 
The network learns to produce similar embeddings for Anchor--Positive pairs and dissimilar embeddings for Anchor--Negative pairs by minimizing the triplet loss. 
This process teaches the model to recognize the underlying kinematic structure of clusters, independent of projection and resolution differences.}
    \label{fig:siam_cnn_diag_training}
\end{figure*} 
\section{ICM velocity maps: {\it XMM-Newton} observations}\label{sec_obs}    

For this study, we used velocity maps of the Virgo \citep{gat22a,gat23b}, Centaurus \citep{gat22b,gat23d}, Ophiuchus \citep{gat23a,gat23e}, and A3266 \citep{gat24a,gat24b} clusters.  
Here we summarize the procedure used to generate the velocity maps. 
The {\it XMM-Newton} EPIC data were reduced with SAS (version 19.1.0) using {\tt epchain}, selecting single-pixel events (PATTERN==0), filtering soft-proton flares (threshold $1.0$~cts/s), and removing bad pixels and CCD edges (FLAG==0). 
Point sources, including central AGN, were detected with {\tt edetect\_chain} (detection likelihood {\tt det\_ml} $>10$) and masked. 
Following \citet{san20} and \citet{gat22a,gat22b}, velocity maps were constructed from adaptive elliptical regions (2:1 axis ratio, tangentially oriented) sized to contain $\sim$500--750 Fe-K counts, sampled on a $0.25$~arcmin grid. 
Spectra from all exposures within each bin were combined and fitted in XSPEC (version 12.11.1) using the cash statistic. 
The ICM emission was modeled with the {\tt apec} model \citep{fos19}, except in Centaurus where a {\tt lognorm} model accounted for its multi-temperature structure; Galactic absorption was included via {\tt tbabs} \citep{wil00}. 
Free parameters included redshift, temperature, metallicity, normalization (and log$\sigma$ for the {\tt lognorm} model), and instrumental background lines (Cu-K$\alpha$, Cu-K$\beta$, Ni-K$\alpha$, Zn-K$\alpha$) were added explicitly. 
This procedure yields line-of-sight velocity precision better than $100$~km/s at the Fe-K line \citep{san20}.

The resulting velocity maps reveal complex and diverse kinematic structures in the ICM across the analyzed clusters.
In the Virgo cluster, the velocity field shows signatures consistent with both AGN-driven outflows and gas sloshing within the core \citep{gat22a}.
For the Centaurus cluster, a clear radial velocity gradient extends to large radii, with distinct variations coincident with cold fronts \citep{gat22b}.
In Ophiuchus, a prominent redshifted--blueshifted interface is detected $\sim150$~kpc east of the cluster center, coinciding with X-ray surface brightness discontinuities \citep{gat23a}.
In A3266, the hot gas exhibits a globally redshifted systemic velocity relative to the cluster mean across the field of view \citep{gat24a}.
Overall, most clusters lack the characteristic spiral or symmetric redshift--blueshift patterns associated with sloshing, suggesting that the line of sight is approximately perpendicular to the sloshing plane \citep{zuh16}. 
Localized high-velocity regions may instead be linked to AGN feedback or merger-induced motions, though weak Fe-K emission in these areas increases velocity uncertainties \citep{gat23a}.

\section{IllustrisTNG}\label{sec_sim} 

The IllustrisTNG project is a suite of state-of-the-art cosmological magnetohydrodynamic simulations that improve upon the original Illustris simulations by incorporating more refined physical models and larger cosmological volumes \citep{spr18,pil18,nai18,mar18,nel18,nel19}. 
Using the moving-mesh code \textsc{AREPO} \citep{spr10}, IllustrisTNG simulates various physical processes including radiative cooling, star formation, supernova and AGN feedback, and chemical enrichment. 
The TNG simulations have been widely applied to study cluster physics, including cool core formation \citep{bar18}, black hole feedback \citep{wei18}, and ram-pressure stripping in dense environments \citep{yun19}. 
The suite comprises TNG50, TNG100, and TNG300, with box sizes of $51.7$, $110.7$, and $302.6~\mathrm{Mpc}$, respectively. 
All simulations adopt a $\Lambda$CDM cosmology consistent with \citet{pla16} and evolve from redshift $z=127$ to $z=0$. 
Structures are identified using the friends-of-friends (FoF) algorithm and the \textsc{SUBFIND} algorithm \citep{spr01,dol09}, and galaxies are defined as subhalos with non-zero stellar mass. 
\begin{figure*}
    \centering 
    \includegraphics[width=0.75\linewidth]{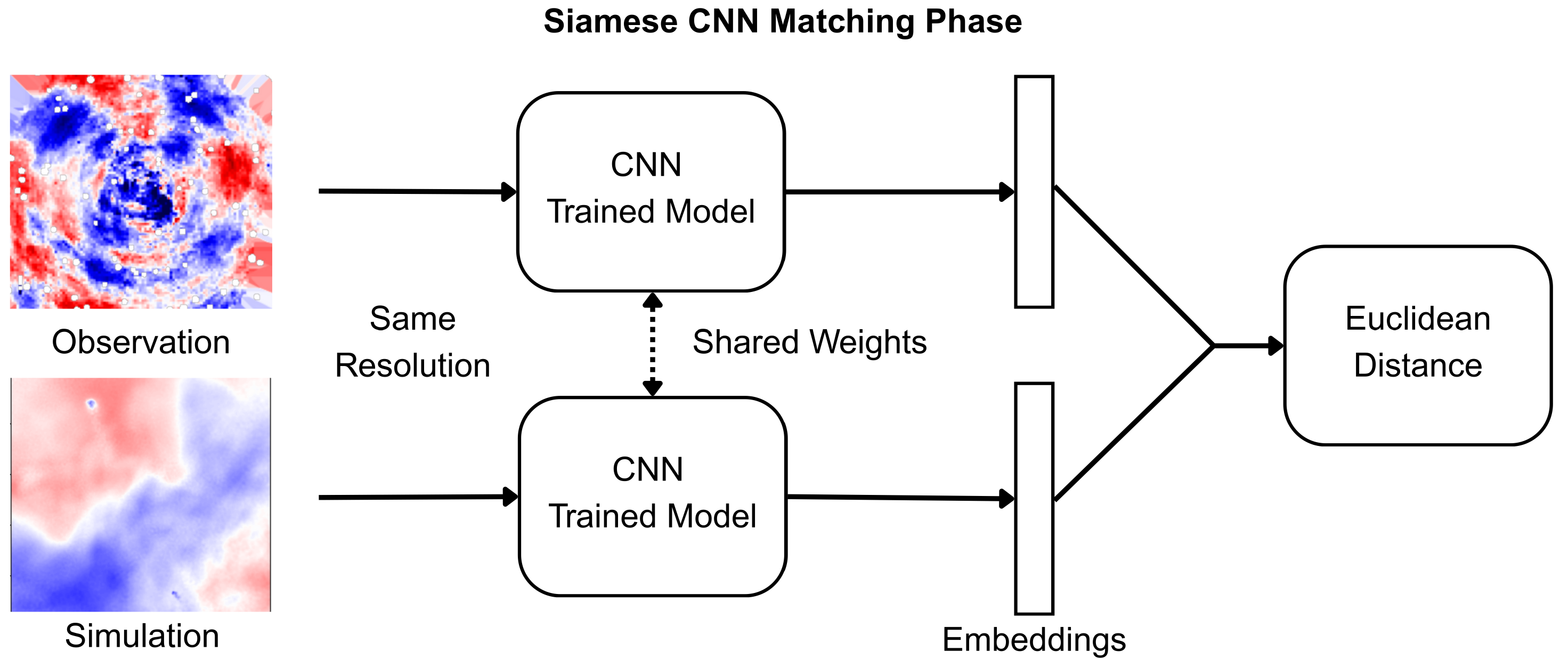} 
    \caption{Diagram of the Siamese CNN matching phase. 
    During matching, the trained Siamese CNN compares each observed {\it XMM-Newton} velocity map to the full library of simulated velocity maps. 
Both the observed map and each simulation are passed through the shared-weight CNN encoder to generate corresponding embedding vectors. 
Similarity between the observation and a given simulation is quantified by the distance between their embeddings in the learned latent space. 
Simulations with the smallest embedding distance are identified as the closest kinematic matches, enabling a quantitative, model-driven comparison between observed cluster dynamics and the cosmological simulations.}
    \label{fig:siam_cnn_diag_matching}
\end{figure*}

We focus on TNG300, which provides the largest cosmological volume and a statistically significant population of massive galaxy clusters. 
TNG300 contains over 280 halos with $\log (M_{200}/M_\odot) > 14$ and several with $\log (M_{200}/M_\odot) > 15$ \citep{soh17}, with $2 \times 2500^3$ resolution elements (dark matter particles and gas cells), a baryon mass resolution of $m_\mathrm{baryon} \sim 1.1 \times 10^7\,M_\odot$, and dark matter particle mass $m_\mathrm{DM} \sim 5.9 \times 10^7\,M_\odot$. 
Galaxy formation histories are traced using SubLink and LHaloTree merger trees \citep{rod15,spr05}, following the main progenitor branch at each timestep. 
The gravitational softening length for the stellar component is $\sim 1~\mathrm{kpc}$ at $z=0$, and the minimum gas cell size can reach $\lesssim 1~\mathrm{kpc}$, enabling detailed modeling of ICM thermodynamics and kinematics. 
From this sample, we selected 40 galaxy clusters at $z=0$ with masses up to $\sim 10^{15}~M_\odot$, consistent with the observed clusters of interest. 

Synthetic observations were generated to enable direct comparison with our {\it XMM-Newton} maps. 
Following \citet{zuh24}, we used the {\tt pyXSIM} package \citep{zuh16b} to produce X-ray emissivity maps. 
Gas cells with $T > 3 \times 10^5$~K were included, while cooler and denser star-forming phases were excluded by applying a density cut of $\rho < 5 \times 10^{-25}$~g~cm$^{-3}$. 
X-ray emissivities (line + continuum) in the 0.1--10~keV band were computed assuming collisional ionization equilibrium with {\tt apec} \citep{fos19}, and projected maps of surface brightness and bulk velocity were constructed from the emissivity-weighted quantities. 
To define the reference velocity frame, the systemic velocity of the Brightest Cluster Galaxy (BCG) was estimated from the halo stellar component. 
The stellar density peak was located, and the mean velocity of stellar particles within an aperture of 20--30~kpc was computed, excluding satellites and intracluster light. 
This velocity vector was adopted as the systemic frame, and all gas and particle velocities were re-centered accordingly before generating synthetic velocity maps in total. 

\begin{table*}
\centering
\small
\caption{Optimized hyperparameters controlling the Siamese CNN architecture.}\label{tab_cnn_hyp}
\begin{tabular}{llc}
\hline 
\textbf{Parameter} & \textbf{Function} & \textbf{Value} \\ 
\hline 
Embedding dimension & Dimensionality of the learned feature space & 64 \\ 
Triplet loss margin & Minimum separation between positive and negative pairs in embedding space & 0.1 \\ 
Learning rate & Step size during optimization & 0.0001 \\ 
Number of epochs & Total training iterations & 100 \\
Batch size & Number of samples per model update & 32 \\
Steps per epoch & Batch iterations per epoch & 75 \\
\hline    
\end{tabular}
\end{table*}

To explore projection effects beyond the Cartesian axes, we generated additional synthetic observations along 101 distinct lines of sight. 
Starting from the standard z-axis projection, we added 100 orientations with polar and azimuthal angles $\theta \in [5^\circ,85^\circ]$ and $\phi \in [0^\circ,175^\circ]$ in $10^\circ$ intervals. 
The directional cosines for each line of sight were computed as:
\begin{align}
x &= \cos{\phi} \cos{\theta},\\
y &= \cos{\phi} \sin{\theta},\\
z &= \sin{\theta}.
\end{align}
These unit vectors enabled construction of projected maps from a broad range of viewing angles. 
The final comparison sample consists of 5016 synthetic velocity maps.

\section{Deep Learning Approach for Velocity Map Matching}\label{sec_ai_stat}

Traditional statistical approaches—such as $\chi^2$ or likelihood-based tests—directly compare binned quantities or predefined features. 
In contrast, a machine learning framework can learn a complex, high-dimensional representation of the data, capturing subtle morphological and kinematic variations that may be overlooked by simpler statistical metrics. 
In this work, we adopt a deep learning approach based on velocity map data to identify the simulated cluster that most closely resembles each {\it XMM-Newton} observation. 
Specifically, we employ a Siamese Convolutional Neural Network (CNN) consisting of two identical branches that share the same architecture and weights, allowing the model to learn a data-driven similarity metric between input maps.

Figure~\ref{fig:siam_cnn_diag_model} illustrates the CNN architecture used in our analysis. 
The network acts as a feature extractor, transforming raw 2D pixel data into a compact numerical vector in a high-dimensional latent space. 
It consists of multiple convolutional layers that progressively extract hierarchical spatial and kinematic features, with max-pooling layers reducing spatial dimensionality while retaining salient features. 
After the final convolutional block, a flattening layer converts the feature maps into a one-dimensional vector, which is then passed through fully connected (dense) layers to refine high-level representations. 
The final dense layer outputs the embedding vector—a concise numerical representation encapsulating the essential kinematic characteristics of the input velocity map. 
This CNN encoder serves as the backbone of the Siamese network.
To ensure fair comparison, the {\it XMM-Newton} velocity maps were re-binned to match the spatial resolution and physical scale of the simulations, using linear interpolation to preserve data integrity.

\begin{figure*}
    \centering 
    \includegraphics[width=0.48\linewidth]{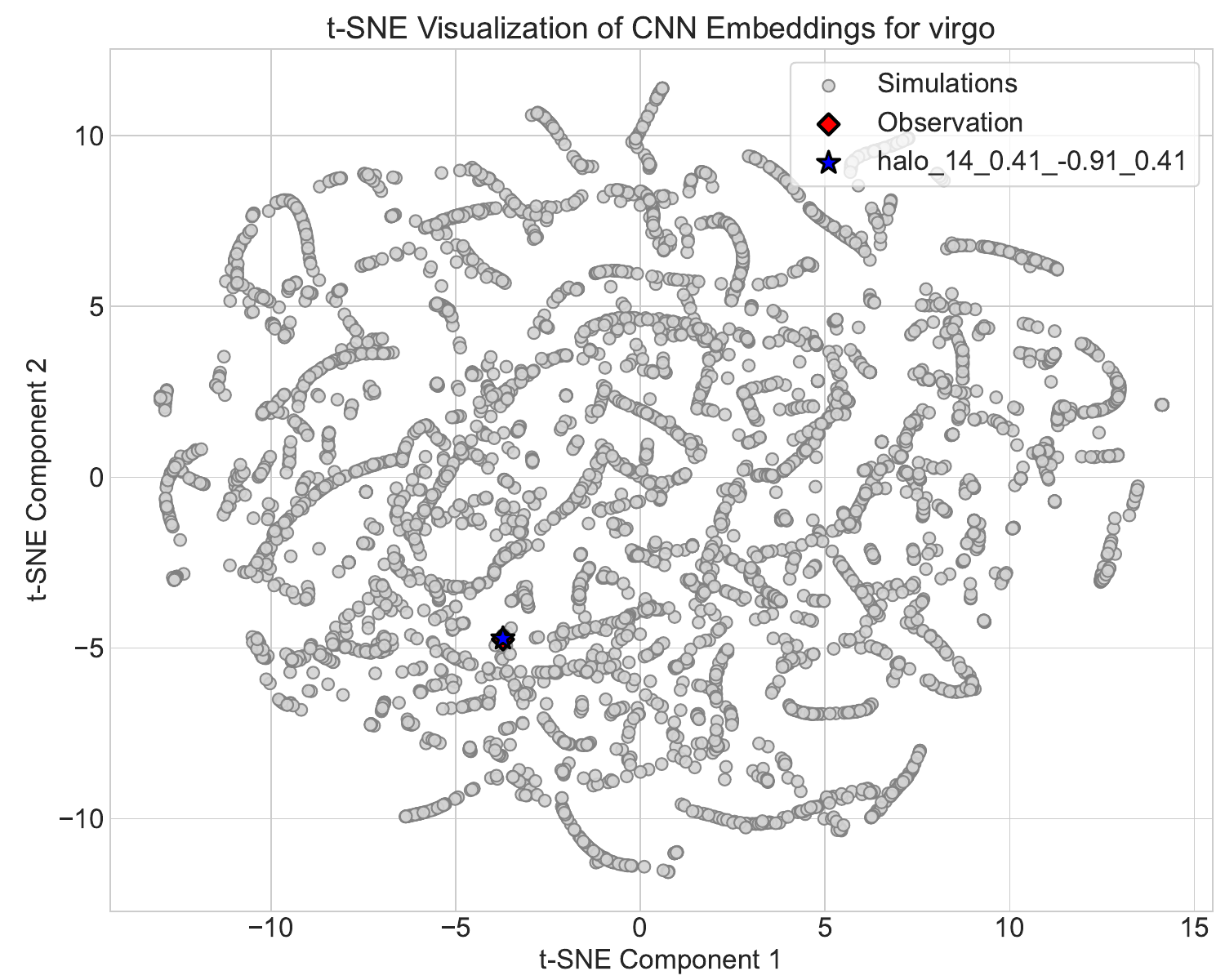} 
    \includegraphics[width=0.48\linewidth]{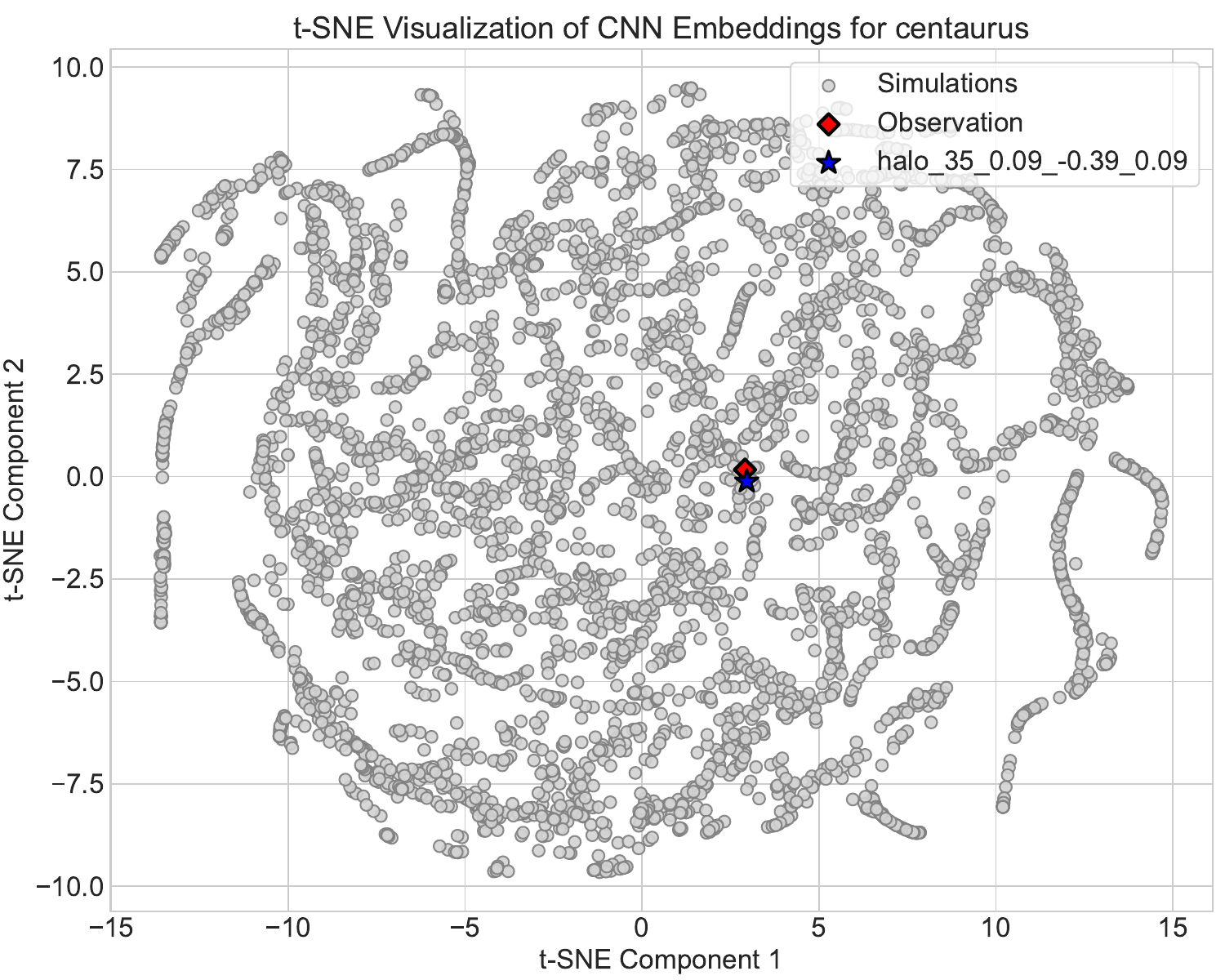} \\
    \includegraphics[width=0.48\linewidth]{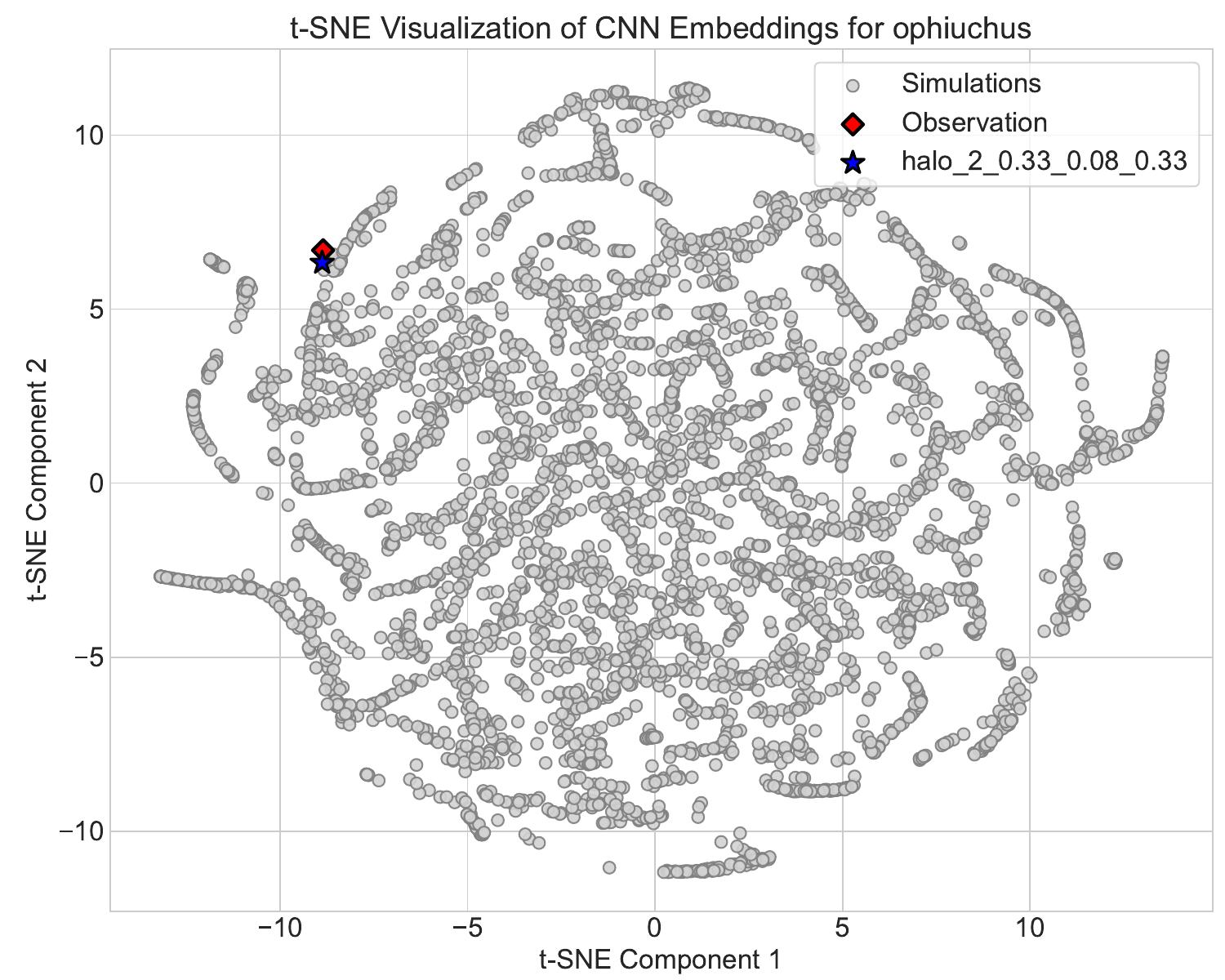} 
    \includegraphics[width=0.48\linewidth]{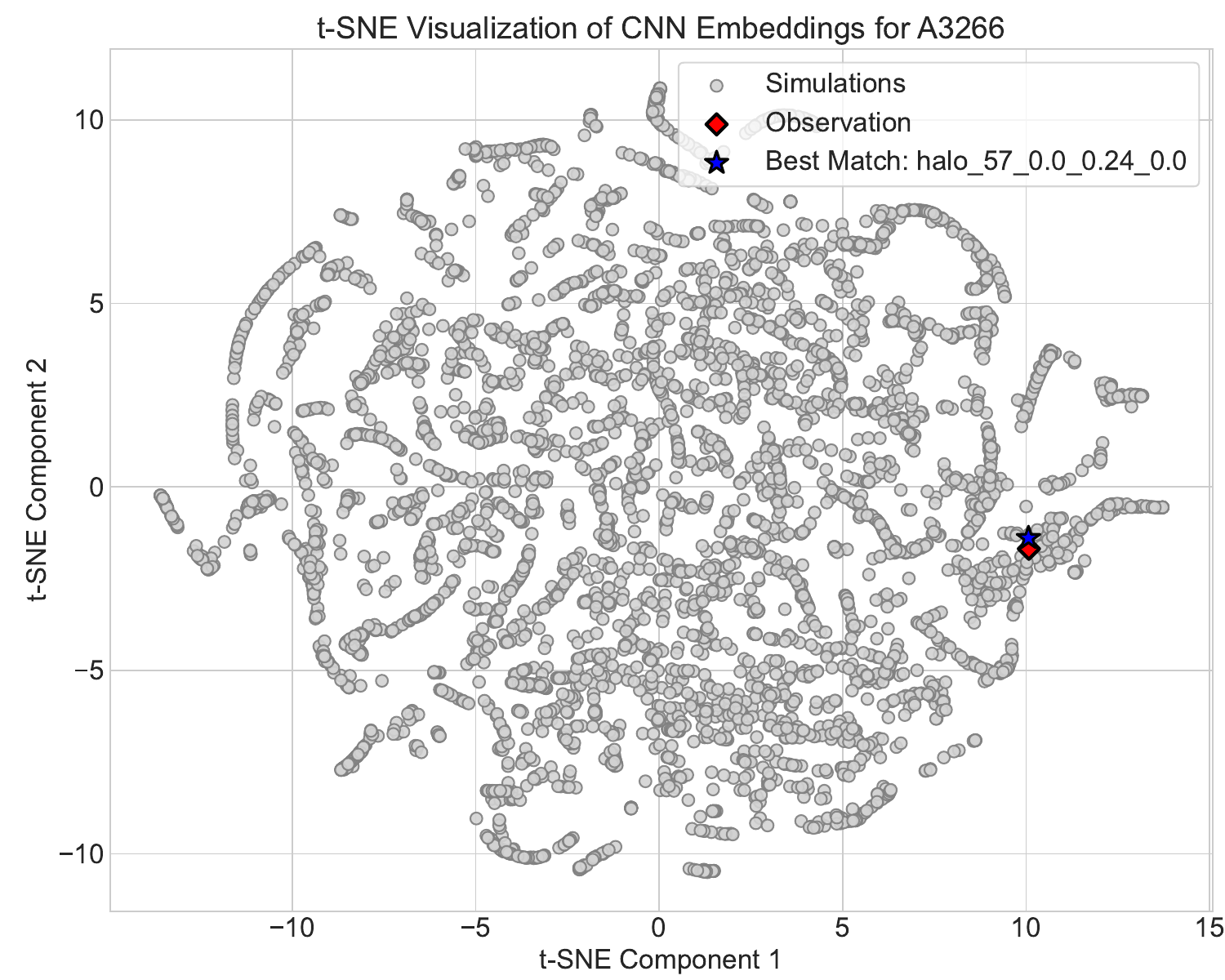} 
    \caption{t-SNE plots comparing the observed velocity maps of four galaxy clusters to a library of TNG simulations. 
    Points represent the similarity of velocity maps, with closer points being more similar. 
    The distribution of TNG simulations is non-uniform, exhibiting curves and filaments corresponding to continuous evolutionary sequences and underlying physical gradients (e.g., merger activity) learned by the network.
    The {\it XMM-Newton} velocity maps are shown in red, and the best-match simulations based on Euclidean distance are highlighted in blue.
    The axes in these t-SNE plots are arbitrary and do not correspond to physical quantities.
    The quantitative similarity ranking is based on the Euclidean distances in the original embedding space.}
    \label{fig:tsne_plots}
\end{figure*}

Figure~\ref{fig:siam_cnn_diag_training} presents the Siamese CNN training process. 
The network is trained using the triplet loss function \citep{sch15}, which operates on three input images: an anchor (A), a positive (P), and a negative (N). 
The loss function is defined as:
\begin{align}
\mathcal{L}_\mathrm{triplet} = \max\big(\|f(A)-f(P)\|_2^2 - \|f(A)-f(N)\|_2^2 + \alpha, 0\big),
\end{align}
where $f$ denotes the CNN embedding function and $\alpha$ is the margin parameter. 
This loss minimizes the distance between anchor and positive embeddings while maximizing the distance between anchor and negative embeddings beyond the margin. 
During training, network weights are updated via back-propagation, effectively organizing the embedding space such that similar velocity maps cluster together and dissimilar maps are pushed apart.

To mitigate overfitting and ensure unbiased performance, the simulated velocity maps were divided into independent training and validation sets. 
The training set comprised 10\% of the total simulations, with tests confirming that increasing this fraction yielded consistent results, demonstrating robustness. 
As additional validation, the network was trained on artificially generated datasets containing the {\it XMM-Newton} observation, perturbed versions, and purely random maps. 
Perturbed images were produced by applying pixel shifts and rotations, while random maps were drawn from a uniform pixel-value distribution. 
The network correctly identified the unperturbed observation as the closest match, confirming that it learned physically meaningful kinematic representations.

Once training was complete, the matching phase began (Figure~\ref{fig:siam_cnn_diag_matching}). 
Here, the trained CNN branch acts as a fixed feature extractor with frozen weights. 
The {\it XMM-Newton} velocity map is first passed through the network to produce its embedding vector. 
Each TNG300 simulation map is processed through the same network to obtain its embedding. 
All simulation maps are resampled to match the resolution of the observation. 
The similarity between an observation and each simulation is quantified via Euclidean distance in embedding space:
\begin{align}
\mathrm{Distance}(O, S_j) = \|f(O) - f(S_j)\|_2 = \sqrt{\sum_{k=1}^{N_\mathrm{emb}} \big[f(O)_k - f(S_j)_k\big]^2}.
\end{align}
The simulation minimizing this distance is identified as the best match, providing a quantitative, data-driven framework for linking simulated halos to observed ICM velocity structures.

\section{CNN-Based Similarity Matching Results}\label{sec_cnn_results}   
\subsection{Hyperparameter Tuning and Model Optimization}
The performance of the Siamese CNN was optimized through a rigorous hyperparameter tuning process. 
Table~\ref{tab_cnn_hyp} lists the parameters governing the model architecture and training dynamics, alongside their optimized values. 
To efficiently constrain these hyperparameters, a random search approach was employed. 
This method systematically explores various parameter combinations within predefined ranges, offering a more efficient discovery of promising regions in the hyperparameter space compared to a fixed grid search. 
During each trial, a new CNN branch and triplet model were initialized and trained. 
The primary objective was to identify the set of hyperparameters that yielded the lowest triplet loss on the training data. 
The weights of the best-performing model were subsequently saved and used for the final embedding generation and similarity matching. 
This systematic optimization ensures a robust network configuration that effectively captures the underlying patterns in the velocity maps. 

\begin{table*}
\centering
\small
\caption{Summary of TNG300 halo properties from the best matches between {\it XMM-Newton} observations and simulations.}\label{tab_tng300_properties}
\begin{tabular}{lcccc}
\hline 
\textbf{Parameter} & Virgo & Centaurus & Ophiuchus & A3266 \\ 
\hline  
Best-match halo & halo\_14 & halo\_35 & halo\_2 & halo\_57 \\ 
Projection $(x,y,z)$ & $(0.41,-0.91,0.41)$ & $(0.09,-0.39,0.09)$ & $(0.33,0.08,0.33)$ & $(0,0.24,0)$ \\
Euclidean distance ($d_\mathrm{best}$) & $0.002$ & $0.002$ & $0.001$ & $0.001$ \\ 
$M_\star(<R_{200})$ [M$_\odot$] & $4.77\times10^{12}$ & $4.12\times10^{12}$ & $1.07\times10^{13}$ & $3.20\times10^{12}$ \\  
$M_\mathrm{gas}(<R_{200})$ [M$_\odot$] & $6.40\times10^{13}$ & $5.49\times10^{13}$ & $1.44\times10^{14}$ & $4.29\times10^{13}$ \\  
$\mathrm{SFR}(<R_{200})$ [M$_\odot$/yr] & $4.59$ & $1.01$ & $1.33$ & $5.49$ \\  
\hline   
\end{tabular}
\end{table*} 

\subsection{Embedding Space Visualization and Analysis}  
To visually inspect the relationships between observed and simulated velocity maps, we employed t-Distributed Stochastic Neighbor Embedding (t-SNE) to reduce the high-dimensional CNN embeddings to a two-dimensional space. 
The results are shown in Figure~\ref{fig:tsne_plots}. 
In this space, similar data points are represented by nearby points, while dissimilar points are farther apart. 
Notably, the simulated clusters are not uniformly scattered; the resulting structures (e.g., filaments and curves) reflect continuous evolutionary sequences and physical gradients captured by the CNN. 
Each panel corresponds to a different galaxy cluster, showing the observed velocity map (red point) and its best match from the TNG300 sample based on Euclidean distance (blue point). 
These plots indicate that the TNG300 sample contains highly analogous objects with velocity structures closely resembling the observed ones. 
It is important to note that while t-SNE preserves local neighborhood structure, the non-linear, non-metric nature of the projection means that distances in 2D do not precisely reflect the true high-dimensional Euclidean distances used for similarity ranking.

To identify not only the single best match but also simulations close in embedded features, we implemented a relative threshold. 
We first determined the Euclidean distance $d_\mathrm{best}$ of the single best-matching simulation to the {\it XMM-Newton} velocity map. 
A simulation was classified as a ``very close'' match if its distance $d_\mathrm{sim}$ satisfied:
\begin{equation}  
d_\mathrm{best} < d_\mathrm{sim} \le d_\mathrm{best} \times (1 + p),
\end{equation} 
where $p$ is a defined percentage threshold (e.g., $p = 0.20$ for a 20\% threshold). 
This approach ensures that the search for similar simulations is contextual and focused around the best-fitting halo. 
Applying this conservative threshold, we find that the ``very close'' matches for each cluster correspond to the same halo from the simulation suite, differing only by their three-dimensional orientation. 
This result validates the Siamese CNN’s capability, demonstrating that the embedding space successfully captures the intrinsic structure of a halo and correctly clusters rotational variants as highly similar rather than treating them as distinct entities.

\section{Kinematic Analysis of the ICM: Observations vs. Simulations}\label{sec_phys}

\begin{figure*}
    \centering 
    \includegraphics[width=0.49\linewidth]{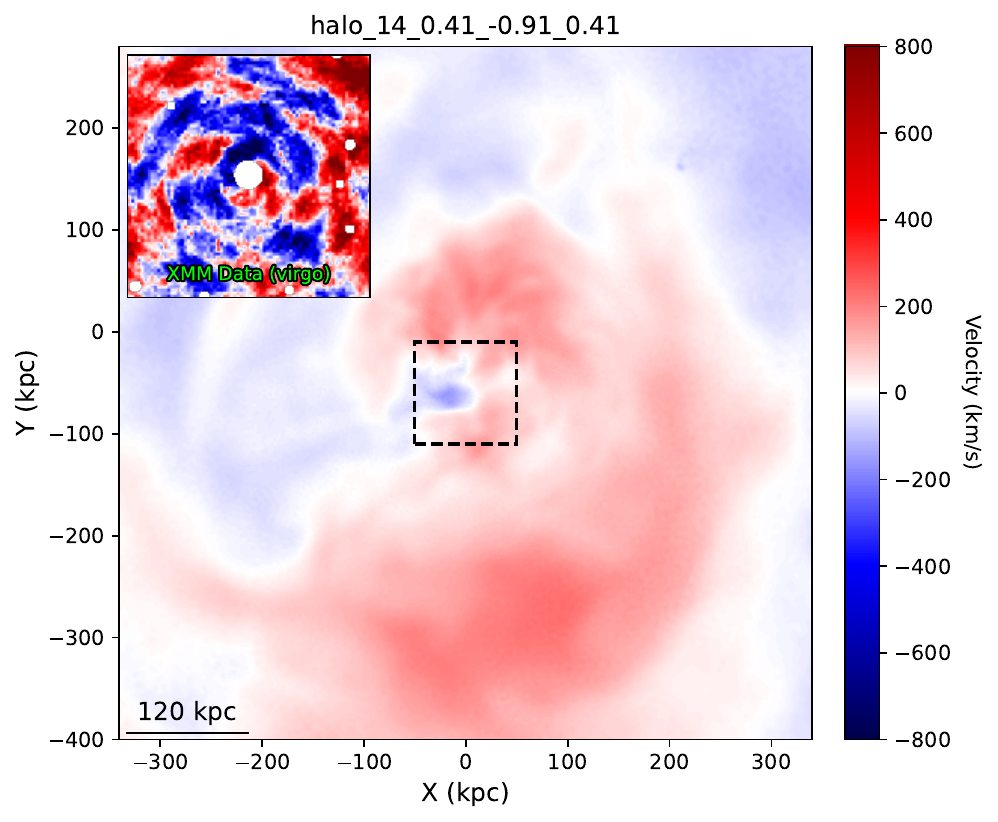}  
    \includegraphics[width=0.49\linewidth]{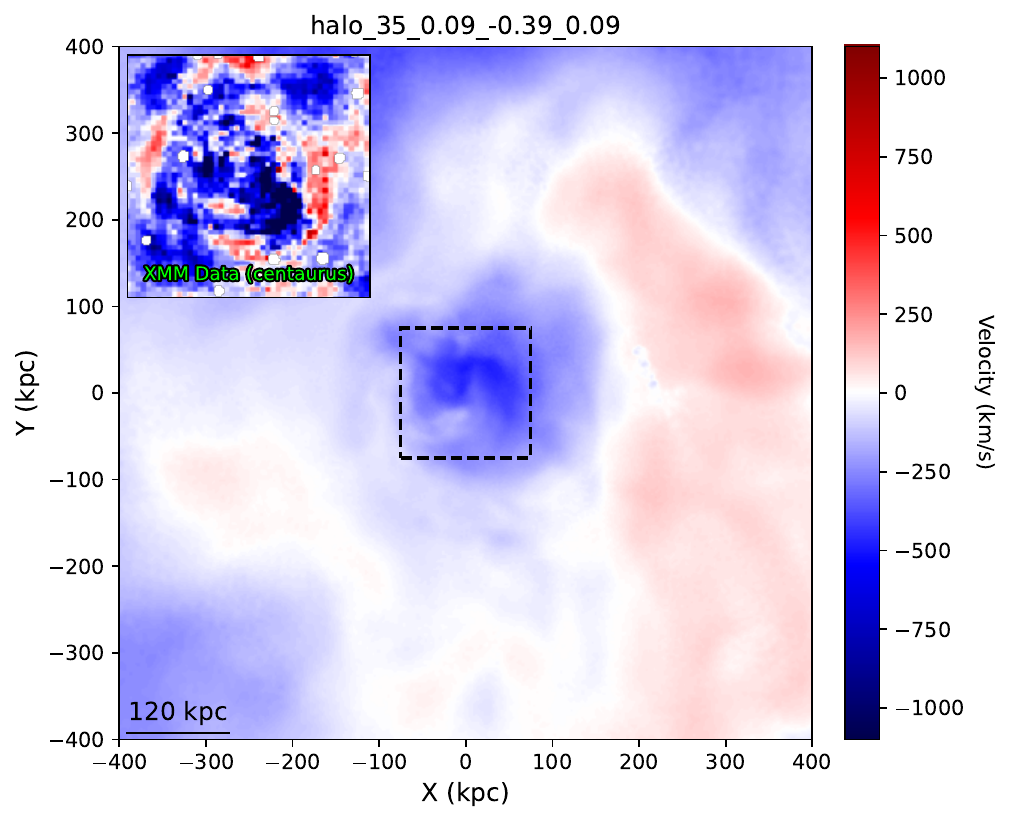}  \\   
    \includegraphics[width=0.49\linewidth]{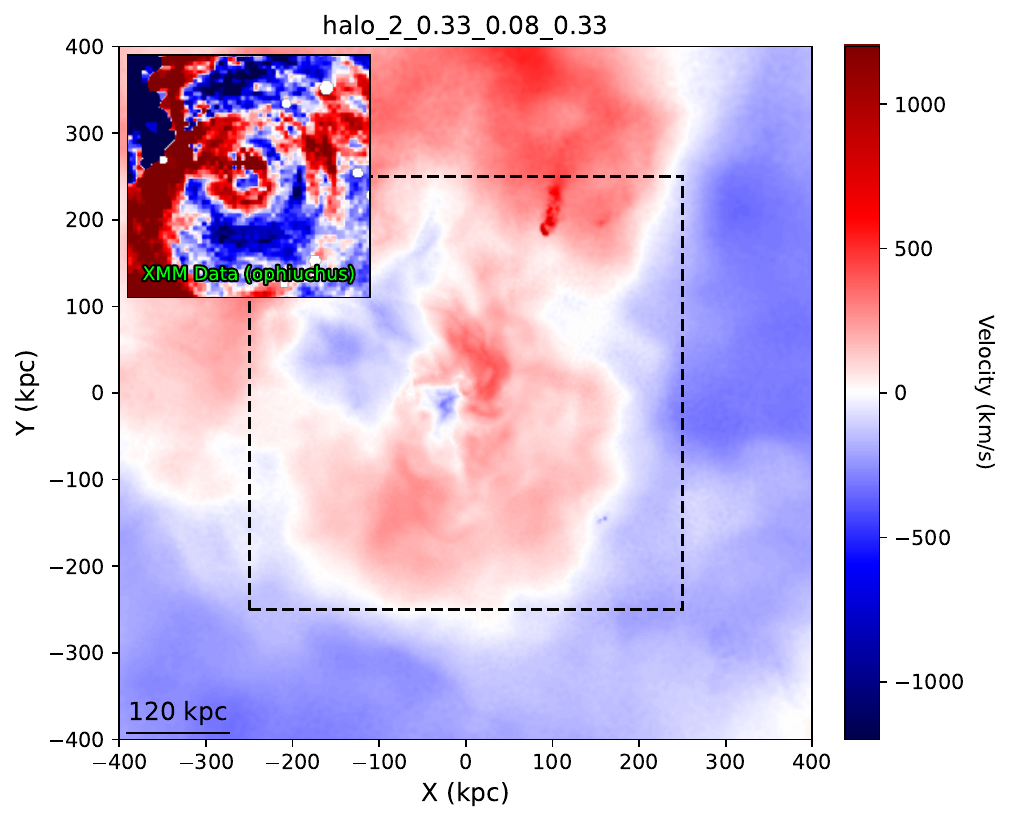}            
    \includegraphics[width=0.49\linewidth]{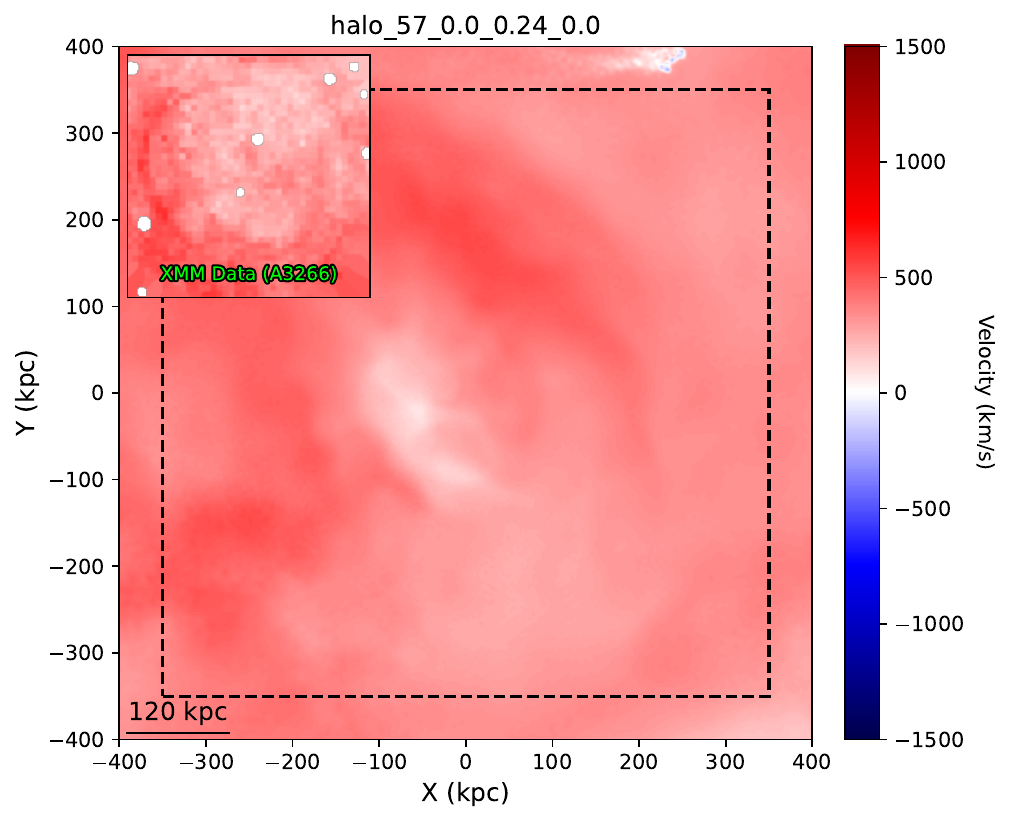}  
    \caption{Best-matching velocity maps from the TNG300 simulation obtained with the Siamese CNN analysis for the {\it XMM-Newton} observations. 
    The zoom-in panel shows the {\it XMM-Newton} data for a region centered on the physical origin (0,0~kpc) of the simulated cluster. 
    These figures illustrate kinematic similarity based on the CNN-learned metric, not pixel-wise visual resemblance.} 
    \label{fig_best_matching_sim_data}
\end{figure*}

Table~\ref{tab_tng300_properties} lists the best-matching halos identified for each observed cluster, including their projected coordinates $(x, y, z)$, the Euclidean distance from the Siamese CNN comparison, and the corresponding total stellar mass, gas mass, and star formation rate within $R_{200}$. 
Figure~\ref{fig_best_matching_sim_data} presents the LOS velocity maps for these best-matching systems. 
In the following subsections, we compare the baryonic properties and velocity structures of the best-matching TNG300 halos with those of the corresponding {\it XMM-Newton} observations.

\subsection{The Virgo Galaxy Cluster}
The Virgo Cluster, centered on M87, exhibits complex ICM kinematics shaped by both merger activity and AGN feedback.
The Siamese CNN identified halo~$\#14$ in TNG300 as the best kinematic match to the observed LOS velocity field, with projected coordinates $(x, y, z) = (0.41, -0.91, 0.41)$ (Figure~\ref{fig_best_matching_sim_data}, top left).
The best-matching simulation reproduces the main baryonic components of Virgo with good fidelity.
At $R_{200}$, the simulated gas mass, $M_{\mathrm{gas}}^{\mathrm{sim}} = 0.64 \times 10^{14}\,M_{\odot}$, falls comfortably within the observational range of $(0.4$–$1.7)\times 10^{14}\,M_{\odot}$ \citep{nul95, sim10, mcc24}, indicating that the TNG300 model accurately captures the global thermodynamic state and gas extent of the cluster.
The stellar mass, $M_{\star}^{\mathrm{sim}} = 4.77 \times 10^{12}\,M_{\odot}$, lies slightly below the lower limit of the observed interval $(5.0$–$20)\times10^{12}\,M_{\odot}$ \citep{fer12, fer16, san19, mor25}, suggesting a mild underestimation of star formation efficiency or past merger-driven buildup in the model.
The total star formation rate (SFR) of $\mathrm{SFR}^{\mathrm{sim}} = 4.6\,M_{\odot}\,\mathrm{yr}^{-1}$ also resides near the lower bound of the observed range ($5$–$20\,M_{\odot}\,\mathrm{yr}^{-1}$) \citep{cor12b, gav12, bos23, edl24}, consistent with the quiescent nature of Virgo.

Both the simulation and the {\it XMM-Newton} data show a large-scale velocity gradient across the cluster core, indicative of gas sloshing triggered by a minor merger \citep{gat22a}.
In the observed map, gas within $r < 30$~kpc is blueshifted relative to M87, while gas at larger radii is redshifted—a pattern reproduced by the best-matching TNG300 halo.
At smaller scales, the {\it XMM-Newton} data show localized, oppositely directed velocity features ($v_{\mathrm{E}} \approx +1200$~km/s, $v_{\mathrm{W}} \approx -1500$~km/s) associated with AGN-driven gas outflows \citep{gat22a}.
While the simulation captures the broad east-west velocity asymmetry, it yields smoother gradients, likely reflecting the limited spatial resolution of the AGN feedback model in TNG300, which lacks explicit jet evolution or resolved bubble dynamics \citep{owe00, for07}.

\subsection{The Centaurus Galaxy Cluster}
The Centaurus Cluster is a cool-core system exhibiting both AGN feedback and complex gas dynamics.
The Siamese CNN identified halo~$\#35$ from TNG300 as the closest kinematic analog, with coordinates $(x, y, z) = (0.09, -0.39, 0.09)$ (Figure~\ref{fig_best_matching_sim_data}, top right).
At $R_{200}$, the simulated gas mass, $M_{\mathrm{gas}}^{\mathrm{sim}} = 5.49 \times 10^{13}\,M_{\odot}$, lies close to the lower edge of the observed range $(6.0$–$15)\times10^{13}\,M_{\odot}$ \citep{san16, wal13b, ver25}, suggesting that the dominant baryonic component is well modeled.
In contrast, the stellar mass is significantly underestimated ($M_{\star}^{\mathrm{sim}} = 4.12 \times 10^{12}\,M_{\odot}$ vs. observed $>1.0\times10^{13}\,M_{\odot}$), likely reflecting either an underproduction of the brightest cluster galaxy and intracluster light or overly strong quenching in massive halos \citep{mis09, sav07, liu12}.
Similarly, the total star formation rate, $\mathrm{SFR}^{\mathrm{sim}} = 1.0\,M_{\odot}\,\mathrm{yr}^{-1}$, is lower than the observed $5$–$15\,M_{\odot}\,\mathrm{yr}^{-1}$ range \citep{mit11, fab16}, consistent with excessive suppression of residual cooling and star formation in the TNG model.

The {\it XMM-Newton} velocity map reveals a prominent blueshifted region southwest of the core, reaching $\sim900$~km/s \citep{gat22b, xri25a}, with no clear spiral pattern characteristic of large-scale sloshing.
The best-matching TNG300 halo reproduces this SW blueshift feature and the absence of a strong sloshing signature, implying that the observed ``Centaurus wind'' could result from bulk ICM motion or weak sloshing along the line of sight, rather than direct AGN jet feedback.
Small-scale redshifted patches observed near the core are less pronounced in the simulation, possibly due to the limited kinematic resolution or observational noise.

\subsection{The Ophiuchus Galaxy Cluster}
The Ophiuchus Cluster is an extremely massive, X-ray luminous system exhibiting disturbed kinematics consistent with a recent or ongoing merger.
The Siamese CNN selected halo~$\#2$ from TNG300 as the best match, with coordinates $(x, y, z) = (0.33, 0.08, 0.33)$ (Figure~\ref{fig_best_matching_sim_data}, bottom left).
The simulated baryonic properties show excellent agreement with observations.
The stellar mass within $R_{200}$, $M_{\star}^{\mathrm{sim}} = 1.07 \times 10^{13}\,M_{\odot}$, aligns with the measured $(1.11^{+0.26}_{-0.43})\times10^{13}\,M_{\odot}$ \citep{dur15}.
The gas mass, $M_{\mathrm{gas}}^{\mathrm{sim}} = 1.44 \times 10^{14}\,M_{\odot}$, also agrees well with the observed range of $(1$–$2)\times10^{14}\,M_{\odot}$ \citep{ang20, gat23a, gat23e}.
The simulated SFR, $\mathrm{SFR}^{\mathrm{sim}} = 1.33\,M_{\odot}\,\mathrm{yr}^{-1}$, slightly exceeds the observational upper limit of $<1\,M_{\odot}\,\mathrm{yr}^{-1}$ \citep{dur15}, suggesting a modest overprediction of residual star formation in an otherwise quenched cool-core cluster.

The observed LOS velocity field shows strong radial asymmetries, with a blueshifted core and a redshifted envelope, separated by a sharp velocity discontinuity $\Delta v \sim 2500$~km/s $\sim150$~kpc east of the center \citep{gat23a, fuj25}.
The best-matching TNG halo reproduces the overall velocity structure—including the asymmetric east-side discontinuity and large-scale velocity amplitudes ($>1000$~km/s)—supporting the interpretation of Ophiuchus as a major off-axis merger.
However, the simulated transition between the blue- and redshifted regions is smoother, consistent with the finite resolution and dissipative numerical treatment of shocks in the simulation.
Nonetheless, the asymmetry between the eastern and western sides of the cluster is well reproduced, indicating that the model captures the merger geometry and resulting large-scale gas dynamics.

\subsection{The A3266 Galaxy Cluster}
Abell~3266 is a dynamically young system exhibiting large-scale bulk motions and substructure consistent with an ongoing major merger \citep{gat24a}.
The Siamese CNN identified halo~$\#57$ from TNG300 as the best kinematic match, with projected coordinates $(x, y, z) = (0.0, 0.24, 0.0)$ (Figure~\ref{fig_best_matching_sim_data}, bottom right).
The best-matching halo shows baryonic properties broadly consistent with observations.
The stellar mass, $M_{\star}^{\mathrm{sim}} = 3.2 \times 10^{12}\,M_{\odot}$, lies within the wide observational range $(0.2$–$12)\times10^{12}\,M_{\odot}$ \citep{gon13, bab14}, while the gas mass, $M_{\mathrm{gas}}^{\mathrm{sim}} = 4.29 \times 10^{13}\,M_{\odot}$, sits at the lower end of the measured $(0.3$–$1.9)\times10^{14}\,M_{\odot}$ range \citep{deg99, hen00, san22}.
The simulated total star formation rate, $\mathrm{SFR}^{\mathrm{sim}} = 5.5\,M_{\odot}\,\mathrm{yr}^{-1}$, remains below the observational upper limit of $\mathrm{SFR}_{\mathrm{tot}} < 10\,M_{\odot}\,\mathrm{yr}^{-1}$ \citep{qui96, hen21}, suggesting a realistic, moderate level of star formation possibly sustained by merger-induced gas compression \citep{hen21}.

The observed {\it XMM-Newton} velocity field shows redshifted gas motions across the entire field, increasing from $\sim300$~km/s in the northwest to $\sim800$~km/s in the southeast, consistent with merger-driven bulk motion \citep{gat24a}.
The TNG300 simulation reproduces this large-scale redshifted gradient and the characteristic C-shaped velocity structure visible in the data, confirming that the observed ICM dynamics likely result from the projected motion of subcluster components during a major merger event \citep{ang15, ang16, ota16, liu18}.

\section{Caveats and Limitations}\label{sec_cav} 
The main caveats in our analysis are as follows: 
\begin{itemize} 
    \item Full {\it XMM-Newton} response modeling was not included in generating the simulated maps due to computational cost and the complexity of multiple pointings. Instead, we compare physical velocity maps directly, a simplification that enables efficient testing of the CNN framework. Future work will explore incorporating more realistic mock observations.  
    \item The initial focus of our machine learning approach has been on the line-of-sight velocity maps. Future work will aim to incorporate a broader range of robust physical observables (including X-ray surface brightness and more accurately constrained temperature and metallicity maps) into the Siamese CNN method, provided such reliable maps can be consistently extracted from the data. 
    \item Our current training and testing paradigm does not explicitly account for the propagation of statistical uncertainties present in the observational or simulated maps. Future implementations will incorporate observational uncertainties to improve network robustness and reliability.  
\end{itemize}

\section{Conclusions}\label{sec_con}
In this work, we developed a deep-learning framework based on a CNN to quantify the morphological similarity between observed and simulated ICM velocity maps. 
This approach enables a fully data-driven, non-parametric comparison between {\it XMM-Newton} observations and TNG300 simulations, providing a robust method for linking observed gas kinematics to cosmological models. 
The Siamese CNN was trained using a triplet-loss objective to construct a discriminative embedding space in which morphologically similar velocity structures lie in close proximity. 
The resulting embeddings successfully cluster different projections of the same simulated halo, demonstrating that the model learns orientation-invariant kinematic features and effectively mitigates projection effects.

A t-SNE visualization of the learned CNN embeddings further shows that each observed cluster is closely associated with a simulated halo in this feature space, indicating that the model captures physically meaningful similarities in ICM velocity morphology. 
The best-matching halos identified for the Virgo, Centaurus, Ophiuchus, and A3266 clusters exhibit velocity structures consistent with those observed by {\it XMM-Newton}, including coherent large-scale gradients and localized substructures linked to AGN feedback or merger-induced motions. 
In agreement with expectations from hydrodynamical simulations, these results suggest that the ICM dynamics in the analyzed systems are shaped by a combination of gas sloshing, AGN-driven outflows, and minor merger activity.

Quantitative comparisons between the observed and simulated baryonic components reveal overall consistency across the sample. 
The best-matching TNG300 halos reproduce the observed gas masses within uncertainties, slightly underpredict the stellar masses, and generally yield star formation rates near or below observed values. 
In the TNG300 framework, AGN feedback becomes increasingly important within radii of approximately $r \lesssim 50$–$100$~kpc, where it begins to compete with or exceed the dynamical impact of sloshing and minor mergers; outside this region, sloshing-induced motions tend to dominate. 
The agreement found here supports the global validity of the TNG300 feedback model while also leaving room for refinement. 
If future work incorporating additional cluster properties (e.g., thermodynamic profiles, radial and azimuthal velocity statistics, and metallicity structure) confirms the systematic trends identified in this study, current assumptions regarding AGN feedback efficiency and star-formation suppression in simulations may require revision.

Overall, the framework presented here provides a scalable and reproducible pathway for connecting high-resolution X-ray observations with large-volume cosmological simulations. 
It establishes a foundation for future studies with next-generation X-ray observatories such as \textit{Athena}, which will deliver unprecedented velocity precision and enable direct tests of ICM kinematics and feedback physics across cosmic time.

\begin{acknowledgements} 
This work was supported by the Deutsche Zentrum f\"ur Luft- und Raumfahrt (DLR) under the Verbundforschung programme (Kartierung der Baryongeschwindigkeit in Galaxienhaufen). 
This work is based on observations obtained with XMM-Newton, an ESA science mission with instruments and contributions directly funded by ESA Member States and NASA. 
This research was carried out on the High Performance Computing resources of the cobra cluster at the Max Planck Computing and Data Facility (MPCDF) in Garching operated by the Max Planck Society (MPG). 
The full simulation outputs, including snapshot data and group catalogs, are publicly available via the \textsc{IllustrisTNG} data release portal\footnote{\url{https://www.tng-project.org}}.
\end{acknowledgements}

\bibliographystyle{aa}
 \newcommand{\noop}[1]{}

\end{document}